\documentclass[12pt]{article}

\newenvironment{wileykeywords}{\textsf{Keywords:}\hspace{\stretch{1}}}{\hspace{\stretch{1}}\rule{1ex}{1ex}}

\usepackage{amsmath,amssymb}
\usepackage{graphicx}
\usepackage{color}
\usepackage{dcolumn}
\usepackage{bm}
\usepackage[numbers,super,comma,sort&compress]{natbib}
\usepackage{hyperref}

\definecolor{background-color}{gray}{0.98}

\title{A boundary-integral approach for the Poisson-Boltzmann equation with polarizable force fields}
\author{Christopher D. Cooper\thanks{Departmento de Ingenier\'ia Mec\'anica and Centro Cient\'ifico Tecnol\'ogico de Valpara\'iso (CCTVal), Universidad T\'ecnica Federico Santa Mar\'ia, Valpara\'iso, Chile}}

\begin{document}

\maketitle

\begin{abstract}
Implicit-solvent models are widely used to study the electrostatics in dissolved biomolecules, which are parameterized using force fields.
Standard force fields treat the charge distribution with point charges, however, other force fields have emerged which offer a more realistic description by considering polarizability. 
In this work, we present the implementation of the polarizable and multipolar force field \texttt{AMOEBA}, in the boundary integral Poisson-Boltzmann solver \texttt{PyGBe}. 
Previous work from other researchers coupled \texttt{AMOEBA} with the finite-difference solver \texttt{APBS}, and found difficulties to effectively transfer the multipolar charge description to the mesh.
A boundary integral formulation treats the charge distribution analytically, overlooking such limitations.
We present verification and validation results of our software, compare it with the implementation on \texttt{APBS}, and assess the efficiency of \texttt{AMOEBA} and classical point-charge force fields in a Poisson-Botlzmann solver.
We found that a boundary integral approach performs similarly to a volumetric method on \texttt{CPU}, however, it presents an important speedup when ported to the \texttt{GPU}.
Moreover, with a boundary element method, the mesh density to correctly resolve the electrostatic potential is the same for stardard point-charge and multipolar force fields.
Finally, we saw that polarizability plays an important role to consider cooperative effects, for example, in binding energy calculations. 
\end{abstract}

\begin{wileykeywords}
Poisson-Boltzmann, Implicit solvent, Polarizable force fields, Boundary element method, Electrostatics
\end{wileykeywords}

\clearpage


\begin{figure}[h]
\centering
\colorbox{background-color}{
\fbox{
\begin{minipage}{1.0\textwidth}
\includegraphics[width=50mm,height=50mm]{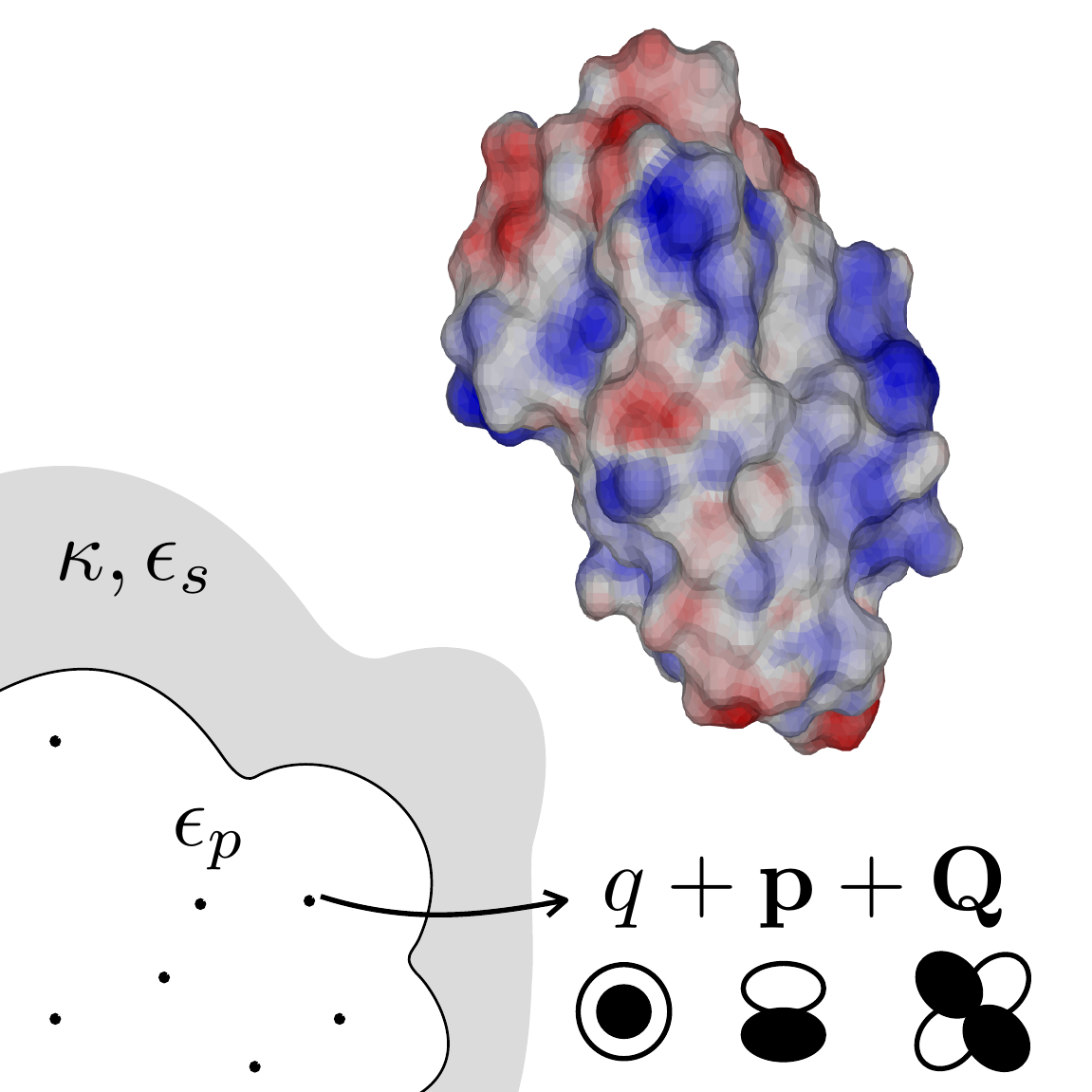} 
\\
Here, we model the electrostatics of biomolecular systems using a continuum approach, while describing the charge distribution inside the molecule with point multipoles that polarize.
In particular, we parameterize the biomolecule with the \texttt{AMOEBA} force field and solve the electrostatic equations with a boundary integral formulation, which integrates the charge distribution analytically.
The implementation is validated, shows good behavior as the size of the biomolecule increases, and is tested for binding energy calculations.
\end{minipage}
}}
\end{figure}

  \makeatletter
  \renewcommand\@biblabel[1]{#1.}
  \makeatother

\bibliographystyle{apsrev}

\renewcommand{\baselinestretch}{1.5}
\normalsize

\clearpage

\section*{\sffamily \Large INTRODUCTION} 



Implicit-solvent models dramatically reduce the problem size in biomolecular simulations.
All of the solvent degrees of freedom are averaged in a continuum dielectric description, whereas the solute is accounted for as a cavity that contains a charge distribution obtained from a force field.
These models are long dated and heavily used by the biophysics/biochemistry community in its various versions,\cite{RouxSimonson1999,Bardhan2012} beginning from Kirkwood's closed expressions for a single spherical cavity,\cite{Kirkwood1934} to Generalized Born models,\cite{StillETal1990} to Poisson and Poisson-Boltzmann solvers.\cite{FogolariETal2002,Baker2004} 
Among the latter, there are numerical packages based on finite difference,\cite{GilsonSharpHonig1987,Bashford1992,GengYuWei2007} finite element,\cite{BakerETal2001} boundary element,\cite{LuETal2006,AltmanBardhanWhiteTidor09,BajajETal2011,GengKrasny2013,CooperBardhanBarba2014} analytical,\cite{FelbergETal2017} and semi-analytical\cite{YapHeadgordon2010} methods, which are usually coupled to point-charge-based force fields.

The accuracy of molecular simulations is greatly influenced by the quality of force fields. 
The most widely used force fields, termed \emph{classical force fields}, describe the charge distribution as a set of point charges, and have proven to be successful in a large number of applications.\cite{PonderCase2003} 
Nevertheless, recently there has been an important development of more elaborate force fields that consider not only the monopole, but also higher order multipole components which may polarize.
This multipole polarizable description yields a more realistic charge distribution, and hence, improved accuracy in the resulting simulations.\cite{PonderCase2003,CieplakETal2009,TongETal2010,CisnerosETal2013,DemerdashYapHeadgordon2014} 
This, however, comes at a higher computational cost.\cite{LippariniETal2014} 
In order to overcome such time limitations, several researchers have coupled polarizable force fields with implicit solvent models. 
For example, there are extensions to Kirkwood's solution\cite{KongPonder1997} and Generalized Born models\cite{SchniedersPonder2007} to account for polarizable multipoles. 
Moreover, we can find several efforts towards including polarizability to Poisson-Boltzmann\cite{MapleETal2005,SchniedersBakerRenPonder2007,TanTanWangLuo2008,TongETal2010,AleksandrovETal2018} and Poisson\cite{LiGordon2007,LippariniBarone2011,LippariniETal2015} solvers.

Recently, there has been an increasing amount of research developing polarizable force fields.\cite{ShiRenSchniedersPiquemal2015} 
Among the most popular ones is \texttt{AMOEBA} (atomic multipole optimized energetics for biomolecular applications),\cite{RenPonder2003,PonderETal2010} which is available in several packages, such as Tinker,\cite{PonderRichards1987,KundrotPonderRichards1991,PappuHartPonder1998} Force Field X,\footnote{\url{https://ffx.biochem.uiowa.edu/}} OpenMM,\cite{FriedrichsETal2009} and Amber.\cite{Case05}
The \texttt{AMOEBA} force field uses point multipoles up to the quadrupole moment to describe the charge distribution, and allows for the dipole moment to have a polarizable component.  
A notable effort in coupling \texttt{AMOEBA} with a Poisson-Boltzmann model is the work by Schnieders and co-workers,\cite{SchniedersBakerRenPonder2007} where the authors extended the widely used \texttt{APBS} solver\cite{BakerETal2001} (using a finite difference method) to account for polarizable multipoles. 
In that work \texttt{AMOEBA} was successfully integrated into \texttt{APBS}, however, some limitations may be found in transferring the charge distribution to the finite difference mesh, due to the high order multipole components (dipole and quadrupole).
The number of mesh nodes to map a point multipole adequately increases with the order of the multipole, to the point that a quadrupole needs 5 evenly-spaced points per dimension.
Moreover, a very fine mesh is required for multipoles near the dielectric interface, to avoid placing charges outside the cavity that represents the biomolecule.

In this work, we present a boundary-integral Poisson-Boltzmann solver, compatible with the \texttt{AMOEBA} force field.
We extended the boundary element method (\texttt{BEM}) code \texttt{PyGBe},\cite{CooperBardhanBarba2014} which uses the formulation presented by Yoon and Lehoff,\cite{YoonLenhoff1990} to account for polarizable multipoles.
In a boundary integral framework, the charge distribution is integrated analytically,\cite{LippariniETal2015} avoiding the limitations encountered in a finite-difference model.
Moreover, the \texttt{BEM} stiffness matrix is not affected by the charge description, making it realtively easy to port into existing boundary-element codes.

\section*{\sffamily \Large METHODOLOGY}

\section*{\sffamily \Large The boundary integral formulation in the implicit solvent model}

The implicit-solvent model divides the domain in a protein ($\Omega_1$) and a solvent ($\Omega_2$) region, interfaced by the solvent excluded surface ($\Gamma$, \texttt{SES}), as sketched in Figure \ref{fig:biomolecule}. 
In the protein region, the dielectric constant is low and there is a charge distribution which is parameterized using force fields.
The solvent is usually water ($\epsilon \approx 80$), with salt.
Continuum electrostatic theory leads to a coupled system to solve for the potential ($\phi$), where the Poisson equation models the protein region and the linearized Poisson-Boltzmann equation governs in the solvent, with appropriate interface conditions.
\begin{align} \label{eq:pde}
\nabla^2 \phi_1(\mathbf{r}) &= - \sum_k \frac{q_k}{\epsilon_1} \delta(\mathbf{r},\mathbf{r}_k) \ \text{ in solute $(\Omega_1)$,}  \nonumber \\ 
\nabla^2\phi_2 (\mathbf{r}) &= \kappa^2 \phi_2(\mathbf{r}) \quad \qquad \ \ \text{ in solvent $(\Omega_2)$,}  \nonumber \\ 
\phi_1 &=\phi_2 \qquad \qquad \qquad \text{ on interface $\Gamma$,}  \nonumber \\ 
\epsilon_1 \frac{\partial \phi_1}{\partial \mathbf{n}} &= \epsilon_2 \frac{\partial \phi_2}{\partial \mathbf{n}}.
\end{align}
Here, the subscripts $1$ and $2$ refer to the protein and solvent regions, respectively, $\kappa$ is the inverse of the Debye length, and $\mathbf{n}$ is a unit normal vector to the \texttt{SES} pointing out of the protein, into the solvent.

If we apply Green's second identity on Equation \eqref{eq:pde}, we get
\begin{align} \label{eq:green_identity}
\phi_{1}+ K_{L}^{\Omega_1}(\phi_{1,\Gamma}) -  V_{L}^{\Omega_1} \left(\frac{\partial}{\partial \mathbf{n}}  \phi_{1,\Gamma}  \right) & = \frac{1}{\epsilon_1} \sum_{k=1}^{N_q}  \frac{q_k}{4\pi|\mathbf{r}_{\Omega_1} - \mathbf{r}_k|}  \quad \text{on $\Omega_1$,} \nonumber \\
\phi_{2} - K_{Y}^{\Omega_2}(\phi_{2,\Gamma}) + V_{Y}^{\Omega_2} \left( \frac{\partial}{\partial \mathbf{n}} \phi_{2,\Gamma} \right) & = 0 \quad \text{on $\Omega_2$,}
\end{align}
where $\phi_{i,\Gamma} = \phi_i(\mathbf{r}_\Gamma)$ is the potential in region $\Omega_i$ computed at the surface ($\Gamma$). $K$ and $V$ are the double- and single-layer potentials, evaluated in $\Omega_i$:
\begin{align} \label{eq:layers}
K_{L/Y}^{\Omega_i}(\phi_{i,\Gamma}) &= \oint_{\Gamma} \frac{\partial}{\partial \mathbf{n}} \left[ G_{L/Y}(\mathbf{r}_{\Omega_i},\mathbf{r}_{\Gamma}) \right]\phi_{i,\Gamma} \, \mathrm{d} \mathbf{r}_\Gamma, \nonumber \\
V_{L/Y}^{\Omega_i} \left( \frac{\partial}{\partial \mathbf{n}} \phi_{i,\Gamma} \right) &= \oint_{\Gamma} \frac{\partial}{\partial \mathbf{n}} \phi_{i,\Gamma} G_{L/Y}(\mathbf{r}_{\Omega_i},\mathbf{r}_{\Gamma})  \, \mathrm{d} \mathbf{r}_\Gamma,
\end{align}
with $G_{L/Y}$ the free space Green's function of the Laplace or linearized Poisson-Boltzmann (Yukawa) equations:
\begin{align} \label{eq:free-space}
G_L(\mathbf{r}_{\Omega_1},\mathbf{r}_{\Gamma}) &= \frac{1}{4\pi|\mathbf{r}_{\Omega_1} - \mathbf{r}_{\Gamma}|}, \nonumber \\
G_Y(\mathbf{r}_{\Omega_2},\mathbf{r}_{\Gamma}) &= \frac{\exp \left( -\kappa |\mathbf{r}_{\Omega_1} - \mathbf{r}_{\Gamma}|\right)}{4\pi|\mathbf{r}_{\Omega_1} - \mathbf{r}_{\Gamma}|}.
\end{align}

Evaluating Equation \eqref{eq:green_identity} at the interface $\Gamma$ gives
\begin{align} \label{eq:integral_eq}
\frac{\phi_{1,\Gamma}}{2}+ K_{L}^{\Gamma}(\phi_{1,\Gamma}) - &V_{L}^{\Gamma} \left(\frac{\partial}{\partial \mathbf{n}}\phi_{1,\Gamma} \right)= \frac{1}{\epsilon_1} \sum_{k=0}^{N_q} \frac{q_k}{4\pi|\mathbf{r}_{\Gamma} - \mathbf{r}_k|} \nonumber \\ 
\frac{\phi_{1,\Gamma}}{2} - K_{Y}^{\Gamma}(\phi_{1,\Gamma}) +  &\frac{\epsilon_1}{\epsilon_2} V_{Y}^{\Gamma} \left( \frac{\partial}{\partial \mathbf{n}} \phi_{1,\Gamma} \right) = 0.
\end{align}
Equation \eqref{eq:integral_eq} is the formulation presented by Yoon and Lenhoff.\cite{YoonLenhoff1990}

\section*{\sffamily \Large The implicit-solvent model with polarizable point multipoles}

To consider a polarizable point-multipole description of the charge distribution inside the protein, the model in Equation \eqref{eq:integral_eq} needs to be extended. 
More specifically, the multipolar description has to be taken into account in the right-hand side of Equation \eqref{eq:integral_eq}, and in the calculation of the electrostatic energy.
Furthermore, the dipole moment in \texttt{AMOEBA} has an induced component, which can be solved with a self-consistent approach,\cite{CramerTruhlar1999} where the electric field and induced dipole are iteratively computed until convergence.

\subsection*{\sffamily \large Computation of the right-hand side}

The right hand side of the Poisson equation in Equation \eqref{eq:integral_eq} is the electrostatic potential due to the charge distribution inside the protein, which is a collection of $N_k$ point charges in a classical force field.
In general, the potential due to a charge distribution $\rho(\mathbf{r})$ is
\begin{equation}\label{eq:pot_charge}
\Phi = \frac{1}{4\pi \epsilon} \int_\Omega \frac{\rho(\mathbf{r}')}{|\mathbf{r} - \mathbf{r}'|} \text{d} \mathbf{r}'.
\end{equation}
Considering a point multipole centered at the origin, up to the quadrupole moment, Equation \eqref{eq:pot_charge} becomes
\begin{equation}\label{eq:pot_multipole}
\Phi(\mathbf{r}) = \frac{1}{4\pi\epsilon} \Bigg[ \frac{1}{|\mathbf{r}|} \underbrace{\int_\Omega \rho(\mathbf{r}')\text{d}\mathbf{r}'}_{q} + \frac{r_i}{|\mathbf{r}|^3} \underbrace{\int_\Omega r'_i\rho(\mathbf{r}') \text{d}\mathbf{r}'}_{p_i} + \frac{r_ir_j}{2|\mathbf{r}|^5} \underbrace{\int_\Omega \rho(\mathbf{r}') (3r'_ir'_j-|\mathbf{r}'|^2\delta_{ij})\text{d}\mathbf{r}'}_{Q_{ij}} \Bigg],
\end{equation}
making the right-hand side of Equation \eqref{eq:integral_eq}:
\begin{equation}\label{eq:rhs_multipole}
\frac{1}{4\pi\epsilon_1} \sum_{k=1}^{N_q} \frac{q_k}{|\mathbf{r}_\Gamma-\mathbf{r}_k|} + p_i \frac{(r_{\Gamma, i}-r_{k,i})}{|\mathbf{r}_\Gamma-\mathbf{r}_k|^3} + Q_{ij}\frac{(r_{\Gamma, i}-r_{k,i})(r_{\Gamma, j}-r_{k,j})}{2|\mathbf{r}_\Gamma-\mathbf{r}_k|^5}
\end{equation}
where $q$ is the monopole (or total charge), $p_i$ the dipole vector, and $Q_{ij}$ the quadrupole tensor of the point multipole.
There are other formulations where the $\frac{1}{2}$ in the quadrupole term is absorbed into $Q_{ij}$.\cite{Stone2013}



\subsection*{\sffamily \large Induced dipole}

The point multipoles of the \texttt{AMOEBA} force field have an induced dipole component.
That is, the force field assigns values of permanent monopole ($q$), dipole ($d_i$), and quadrupole ($Q_{ij}$) moments, and a polarizability ($\alpha$), such that there is an induced dipole ($\boldsymbol{\mu}$)
\begin{equation}\label{eq:induced_dipole}
\boldsymbol{\mu} = \alpha \mathbf{E},
\end{equation}
where $\mathbf{E}=-\nabla \phi$ is the electric field at the location of the point multipole. 
The total dipole moment in Equation \eqref{eq:rhs_multipole} is then
\begin{equation}\label{eq:dipole_decomposition}
\mathbf{p} = \boldsymbol{\mu} + \mathbf{d}.
\end{equation}

\subsubsection*{\sffamily \normalsize Calculation of the electric field}
To compute the induced dipole moment in Equation \eqref{eq:induced_dipole}, it is easier to treat the electrostatic potential $\phi$ (hence, the electric field) as two separate components: one from the reaction of the solvent ($\phi_\text{solvent}$), and from the multipoles ($\phi_\text{multipoles}$): 
\begin{equation} \label{eq:pot_comp}
\phi = \phi_\text{solvent} + \phi_\text{multipoles}.
\end{equation}

We can obtain $\phi_\text{solvent}$ by substracting out the influence of the multipoles to Equation \eqref{eq:green_identity}, which leaves:
\begin{equation} \label{eq:pot_solvent}
\phi_\text{solvent} =  -K_{L}^{\Omega_1}(\phi_{1,\Gamma}) +  V_{L}^{\Omega_1} \left(\frac{\partial}{\partial \mathbf{n}}  \phi_{1,\Gamma}  \right).
\end{equation}
The $i^{th}$ component of the electric field $E^\text{solvent}_i=-\partial \phi_\text{solvent}/\partial r_i$ is then
\begin{equation} \label{eq:Efield_solv}
E^\text{solvent}_i=-\frac{\partial \phi_\text{solvent}}{\partial r_i} = -\int_\Gamma \frac{\partial}{\partial r_i} G_L(\mathbf{r},\mathbf{r}') \frac{\partial \phi_1}{\partial \mathbf{n}'}(\mathbf{r}') \mathrm{d}\mathbf{r}' + \int_\Gamma \frac{\partial}{\partial r_i} \frac{\partial G_L}{\partial \mathbf{n}'}(\mathbf{r},\mathbf{r}')\phi_1(\mathbf{r}') \mathrm{d}\mathbf{r}',
\end{equation}
where the derivatives of the Green's function are detailed in Equation \eqref{eq:Efield_solv_der}.

On the other hand, the electric field due to $N_m$ multipoles at the location of multipole $l$ is the gradient of the Coulombic-type electrostatic potential from Equation \eqref{eq:pot_multipole}, taken at the evaluation point.
This is
\begin{align} \label{eq:Efield_mult}
E_k^\text{mult} = -\frac{\partial}{\partial r^{(l)}_k} \phi_\text{mult}(\mathbf{r}^{(l)}) =& -\sum_{m=1, m \neq l}^{N_m} \frac{1}{4\pi\epsilon} \frac{\partial}{\partial r^{(l)}_k}\left[ \frac{1}{|\mathbf{r}^{(l)}-\mathbf{r}^{(m)}|} q^{(m)}\right. \nonumber\\
        &\left.+ \frac{r_i^{(l)}-r_i^{(m)}}{|\mathbf{r}^{(l)}-\mathbf{r}^{(m)}|^3} p^{(m)}_i + \frac{(r_i^{(l)}-r_{i}^{(m)})(r_{j}^{(l)}-r_{j}^{(m)})}{2|\mathbf{r}^{(l)}-\mathbf{r}^{(m)}|^5} Q^{(m)}_{ij} \right].
\end{align}
The detailed expressions for the derivatives of each term are shown in Equation \eqref{eq:derivatives_Emult}.

\texttt{AMOEBA} uses a group-based polarization scheme for permanent multipoles\cite{RenPonder2002} where, in this case, multipoles of the same group do not polarize each other.
Then, we need to consider a masking rule in Equation \eqref{eq:Efield_mult} that zeroes out the contribution of permanent multipole $m$ when it is in the same polarization group as multipole $l$.
Moreover, the field generated by induced dipoles ($\boldsymbol{\mu}$ in Equation \eqref{eq:dipole_decomposition}) is damped with a Thole-like scheme.\cite{Thole1981,RenPonder2003}

Having both components of the electric field from equations \eqref{eq:Efield_mult} and \eqref{eq:Efield_solv}, we can compute the total electric field at the location of the point multipoles as
\begin{equation} \label{eq:Efield_tot}
\mathbf{E} = \mathbf{E}^\text{solvent}+\mathbf{E}^\text{mult},
\end{equation}
which goes into Equation \eqref{eq:induced_dipole} to obtain the induced dipole moment.

\subsubsection*{\sffamily \normalsize Self-consistent induced field}
The dipole moment is an input to the implicit-solvent model, however, the electric field has an influence in the induced dipole component.
Then, we need to solve for the induced dipole with a self-consistent scheme, which is summarized below:
\begin{enumerate}
\item Guess $\boldsymbol{\mu}$ (for example, $\boldsymbol{\mu}=0$).
\item Compute $\phi$ and $\partial \phi/\partial \mathbf{n}$ on $\Gamma$ with Equation \eqref{eq:integral_eq}.
\item Find $\mathbf{E}^\text{solvent}$ and $\mathbf{E}^\text{mult}$ with Equation \eqref{eq:Efield_solv} and Equation \eqref{eq:Efield_mult}.
\item Calculate the total electric field with \eqref{eq:Efield_tot}.
\item Get $\boldsymbol{\mu}$ with Equation \eqref{eq:induced_dipole}.
\item Go back to step 2 with the new $\boldsymbol{\mu}$, until convergence.
\end{enumerate}
We update the induced dipole with a successive-over relaxation (\texttt{SOR}) scheme, using a coefficient of $\omega=0.7$.
After reaching convergence of the induced dipole, we can move on to compute the solvation energy.

\subsection*{\sffamily \large Calculation of the solvation energy}

The solvation free energy ($\Delta G$) is the energy required to dissolve a molecule. 
In other words, it is the difference in free energy between the molecule in vacuum state and in the solvent, plus the energy spent polarizing the multipoles:
\begin{equation}\label{eq:energy_general}
\Delta G = G_\text{diss} - G_\text{vac} + W_\text{pol}.
\end{equation}
Assuming a linear dielectric, we can calculate the free energy as
\begin{equation} \label{eq:energy_general}
G = \frac{1}{2}\int_\Omega \rho (\mathbf{r}) \phi(\mathbf{r}) \mathrm{d}\mathbf{r},
\end{equation}
where $\phi$ is the potential and $\rho$ the charge distribution.
Then, we can write the solvation energy as
\begin{equation}\label{eq:deltaG_aux}
\Delta G = \frac{1}{2}\int_\Omega \rho_\text{diss}(\mathbf{r})\phi_\text{diss}(\mathbf{r})\mathrm{d}\mathbf{r} - \frac{1}{2}\int_\Omega \rho_\text{vac}(\mathbf{r})\phi_\text{vac}(\mathbf{r})\mathrm{d}\mathbf{r}+ W_\text{pol}
\end{equation}
Decomposing the potential in dissolved state into \emph{solvent} and \emph{Coulombic} components, as in Equation \eqref{eq:pot_comp}, Equation \eqref{eq:deltaG_aux} becomes
\begin{equation}\label{eq:deltaG_aux2}
\Delta G = \frac{1}{2}\int_\Omega \rho_\text{diss}(\mathbf{r})\phi_\text{solv}(\mathbf{r})\mathrm{d}\mathbf{r} + \frac{1}{2}\int_\Omega \rho_\text{diss}(\mathbf{r})\phi_\text{mult}(\mathbf{r})\mathrm{d}\mathbf{r} - \frac{1}{2}\int_\Omega \rho_\text{vac}(\mathbf{r})\phi_\text{vac}(\mathbf{r})\mathrm{d}\mathbf{r}+ W_\text{pol}
\end{equation}

In standard non-polarizable force fields, the charge distribution does not have an induced component, hence $\rho_\text{diss}=\rho_\text{vac}$ and $\phi_\text{mult} = \phi_\text{vac}$, and only the first term in the right-hand side survives.
However, in the case of polarizable force-fields, like \texttt{AMOEBA}, the induced dipole is different in the solvated and vacuum states, and the second and third terms on the right hand side of Equation \eqref{eq:deltaG_aux2} do not cancel out.

In the context of this work, the charge distribution $\rho$ is a collection of point multipoles, and the integral becomes a sum.
Then, for the case of $N_m$ point multipoles, the energy in Equation \eqref{eq:energy_general} becomes
\begin{equation}\label{eq:energy_multipole}
G = \frac{1}{2}\sum_{m=1}^{N_m} q^{(m)} \phi (\mathbf{r}^{(m)}) + p_i^{(m)} \frac{\partial}{\partial r_i} \phi(\mathbf{r}^{(m)}) + \frac{1}{6} Q^{(m)}_{ij} \frac{\partial}{\partial r_i}\frac{\partial}{\partial r_j} \phi(\mathbf{r}^{(m)}).
\end{equation}
If $\phi=\phi_\text{solv}$ and $p_i=p_{i,\text{dissolve}}$ corresponds to the total dipole in dissolved state, Equation \eqref{eq:energy_multipole} gives the solvent contribution to energy (first term in Equation \eqref{eq:deltaG_aux2}).
Then, if $\phi=\phi_\text{mult}$ and $p_i=p_{i,\text{dissolve}}$, Equation \eqref{eq:energy_multipole} yields the coulombic energy due to the point multipoles in dissolved state (second term in Equation \eqref{eq:deltaG_aux2}). Finally, if $\phi=\phi_\text{vac}$ and $p_i = p_{i,\text{vacuum}}$, one obtains the coulombic energy in vacuum (second-to-last term in Equation \eqref{eq:deltaG_aux2}).
In Equation \eqref{eq:energy_multipole}, $\frac{1}{6}$ becomes $\frac{1}{3}$ when the formulation considers the $\frac{1}{2}$ term of Equation \eqref{eq:rhs_multipole} inside $Q_{ij}$.

We already derived the expressions for $\phi_\text{solv}$ in Equation \eqref{eq:pot_solvent} and $\partial \phi_\text{solv}/\partial r_i$ in Equation \eqref{eq:Efield_solv}. 
The second derivative of $\phi_\text{solv}$ that multiplies the quadrupole in Equation \eqref{eq:energy_multipole} is 
\begin{equation} \label{eq:ddphi}
\frac{\partial}{\partial r_k}\frac{\partial}{\partial r_i}\phi_\text{solv} = \int_\Gamma \frac{\partial}{\partial r_k}\frac{\partial}{\partial r_i} G_L(\mathbf{r},\mathbf{r}') \frac{\partial \phi_1}{\partial \mathbf{n}}(\mathbf{r}') \mathrm{d}\mathbf{r}' - \int_\Gamma \frac{\partial}{\partial r_k}\frac{\partial}{\partial r_i} \frac{\partial G_L}{\partial \mathbf{n}}(\mathbf{r},\mathbf{r}')\phi_1(\mathbf{r}') \mathrm{d}\mathbf{r}',
\end{equation}
where the required derivatives of the Green's function are detailed in Equation \eqref{eq:sec_derivatives}.

The Coulomb potential from a collection of point multipoles is shown in Equation \eqref{eq:pot_multipole} and its first derivative in Equation \eqref{eq:Efield_mult}.
To compute the energy in Equation \eqref{eq:energy_multipole}, the second derivative of this potential is also required for the quadrupole component, and its terms are explicited in Equation \eqref{eq:second_derivative}.

The last term in Equation \eqref{eq:energy_general} is known as the polarization energy, and corresponds to the amount of energy required to polarize the multipoles. 
B\"ottcher \cite{Bottcher1973} derived an expression for $W_\text{pol}$ through the charging process of a spherical cavity by a single polarizable dipole, arriving to:
\begin{equation}
W^\text{single}_\text{pol} = \frac{1}{2}\boldsymbol{\mu}\cdot\mathbf{E}_\text{pol},
\end{equation}
where $\boldsymbol{\mu}$ is the induced dipole component and $\mathbf{E}_\text{pol}$ the total electric field that polarizes the multipole.
In that derivation, the dipole was polarized starting from $\boldsymbol{\mu}=0$, however, in our case, there are several multipoles that are already polarized in the original (vacuum) state.
Then, the total polarization energy is
\begin{align}\label{eq:pol_energy}
W_\text{pol} &= \frac{1}{2} \sum_{m=1}^{N_m} \boldsymbol{\mu}^{(m)}\cdot\mathbf{E}_\text{pol}(\mathbf{r}^{(m)}) - \boldsymbol{\mu}^{v(m)}\cdot\mathbf{E}^v_\text{pol}(\mathbf{r}^{(m)}) \nonumber\\
&= -\frac{1}{2} \sum_{m=1}^{N_m} \mu_i^{(m)} \frac{\partial}{\partial r_i}\phi(\mathbf{r}^{(m)}) - \mu_i^{v(m)} \frac{\partial}{\partial r_i}\phi^{v}(\mathbf{r}^{(m)})
\end{align}
where the superscript $v$ denotes the vacuum state.
Comparing Equation \eqref{eq:pol_energy} with Equation \eqref{eq:energy_multipole}, we realize that the polarization energy cancels out with the induced dipole (as $p_i = \mu_i + d_i$), and we can replace $p_i$ with $d_i$ to obtain
\begin{equation}\label{eq:energy_multipole_polarizable}
G = \frac{1}{2}\sum_{m=1}^{N_m} q^{(m)} \phi (\mathbf{r}^{(m)}) + d_i^{(m)} \frac{\partial}{\partial r_i} \phi(\mathbf{r}^{(m)}) + \frac{1}{6} Q^{(m)}_{ij} \frac{\partial}{\partial r_i}\frac{\partial}{\partial r_j} \phi(\mathbf{r}^{(m)}).
\end{equation}
If we use Equation \eqref{eq:energy_multipole_polarizable}, rather than \eqref{eq:energy_multipole}, the polarization energy is already being considered, and there is no need to explicitly calculate it.

The permanent multipoles for the calculation of $\phi_\text{mult}$ and $\phi_\text{vac}$ and their derivatives are not masked with \texttt{AMOEBA}'s group-based scheme in the energy calculation. 
However, the induced dipole component is damped by the same Thole-like model as the electric field calculation in Equation \eqref{eq:Efield_mult}.

\subsection*{\sffamily \large Calculation of the binding energy}

We can compute the electrostatic component of the energy of binding between two dissolved biomolecules according to the thermodynamic cycle detailed in Figure \ref{fig:thermo_cycle}.
For biomolecules `A' and `B' complexed in `AB', the binding energy in dissolved state is
\begin{equation} \label{eq:bind_energy}
\Delta G_\text{bind,diss} = \Delta G_\text{solv,AB} + \Delta G_\text{bind,vac} - \Delta G_\text{solv,A} - \Delta G_\text{solv,B}
\end{equation}
where $\Delta G_\text{solv,(A,B,AB)}$ is the solvation energy of molecule `A', `B', or the complex `AB', computed with Equation \eqref{eq:energy_general}.
On the other hand, $\Delta G_\text{bind,vac}$ is the binding energy in vacuum, which is the energetic difference between bound and unbound states, plus the energy required to polarize the multipoles:
\begin{equation}\label{eq:bind_vac}
\Delta G_\text{bind,vac} = G_\text{AB} - G_\text{A} - G_\text{B} + W_\text{pol}.
\end{equation}
Here, $G$ is the coulombic energy in vacuum state for each case, computed with Equation \eqref{eq:energy_multipole}.
Note that in the case of standard point-charge force fields, there is no polarization energy ($W_\text{pol}$), and the coulombic energy is the same in vacuum and dissolved states.
To compute $\Delta G_\text{bind,vac}$, we can use Equation \eqref{eq:energy_multipole_polarizable} to consider the polarization energy implicitly.

From the thermodynamic cycle in Figure \ref{fig:thermo_cycle} we can also compute the relative binding energy, given by the difference in binding energy between the dissolved and vacuum states:
\begin{equation} \label{eq:rel_bind_energy}
\Delta \Delta G_\text{bind} = \Delta G_\text{bind,diss} - \Delta G_\text{bind,vac} = \Delta G_\text{solv,AB} - \Delta G_\text{solv,A} - \Delta G_\text{solv,B}.
\end{equation}

\section*{\sffamily \Large Algorithmic and computational details}
\subsection*{\sffamily \large The boundary element method.}
We use a boundary element method (\texttt{BEM}) to solve the system in Equation \eqref{eq:integral_eq} numerically.
In it, we discretize the molecular surface in flat triangular panels, assume a piecewise constant distribution of the potential and its derivative, and use collocation to generate a linear system of equations, which we solve with a generalized minimum residual (\texttt{GMRES}) method.\cite{SaadSchultz1986}

Integrals are calculated with Gaussian quadrature, however, due to the $1/r$ nature of the Green's function, there are singular integrals that are difficult to solve numerically.
For this reason, depending on the distance between the collocation node and the integration panel, we distiguish three integration regions: singular, near-singular, and far-away.
The integral becomes singular when the collocation point is inside the integration panel, and we use a semi-analytical approach\cite{HessSmith1967,ZhuHuangSongWhite2001} that places Gauss  nodes on the edges of the element.
If the collocation point and the integration panel are close-by, the integrand is nearly singular, and we use high order quadrature rules.
For example, we commonly place a threshold $2\sqrt{A}$ away from the collocation point (where $A$ is the area of the element containing the collocation point) and use $19$ Gauss nodes to compute the integrals within that distance.
Finally, beyond this threshold the integrand is smooth enough that we can use a low order approximation with $1$, $3$, or $4$ Gauss nodes. 

\subsection*{\sffamily \large The treecode algorithm}
The most time consuming part of the solver is a matrix-vector product (an $\mathcal{O}(N^2)$ process) inside the \texttt{GMRES}, done once in every iteration.
As we are using collocation and Gaussian quadrature with free-space Green's functions, the matrix-vector product can be seen as a N-body problem where the sources of mass are the Gauss nodes, and we evaluate the Green's function at the collocation points.
There are several ways to accelerate N-body calculations to $\mathcal{O}(N\log N)$, and even $\mathcal{O}(N)$, for example, FFT-based methods,\cite{PhillipsWhite1997,AltmanETal2006} fast multipole method,\cite{GreengardRokhlin1987} and treecode.\cite{BarnesHut1986}
In this work, we use the treecode algorithm, which has been revised for the Green's functions of the Laplace\cite{LindsayKrasny2001} and linearized Poisson-Boltzmann equations.\cite{LiJohnstonKrasny2009}

The treecode clusters \emph{sources} (in this case, Gauss nodes) and \emph{targets} (collocation points) in boxes of an octree structure, where the boxes in the lowest level of the tree contain less than a critical number of particles.
Then, if a box is far enough from a target, the influence of the sources in that box is approximated using a Taylor expansion of order $P$ around the center of the box.
If the box is not far enough, the algorithm checks for child boxes and performs the same operation, until a lowest-level box is reached, and the source-target interaction is performed directly. 
To determine if a box is far enough from a target, we use the multipole-acceptance criterion (\texttt{MAC}), which is:
\begin{equation}
\theta > \frac{R_b}{R_{tb}}
\end{equation}
where $R_b$ is the size of the box and $R_{tb}$ the box-target distance.
This reduces the computational complexity from $\mathcal{O}(N^2)$ to $\mathcal{O}(N\log N)$, allowing us to control accuracy with the order of the Taylor expansion ($P$) and the value of $\theta$ in the multipole-acceptance criterion.
Further details on the implementation of the treecode in the boundary element method can be found elsewhere.\cite{GengKrasny2013,CooperBarba-share154331}

The most time consuming parts of the algorithm are the source-target and box-target interactions, which are completely target independent.
For this reason, the treecode is highly parallelizable, and maps very well to the architecture of the \texttt{GPU}. 
\texttt{PyGBe} uses \texttt{GPU} acceleration, via \texttt{PyCUDA}, for the computation of the source-target and box-target interactions, and also for the computation of the right-hand side in Equation \eqref{eq:rhs_multipole} and the energy in Equation \eqref{eq:energy_multipole_polarizable}.




\section*{\sffamily \Large RESULTS}
We implemented the method described above in a forked Github repository\footnote{\url{https://github.com/barbagroup/pygbe}}\footnote{\url{https://github.com/cdcooper84/pygbe}} of the open-source code \texttt{PyGBe}.\cite{CooperBardhanBarba2014,CooperETal2016} 
The results presented in this section were obtained on a workstation with two \texttt{Intel Xeon E5-2680v3} \texttt{CPU}s and one \texttt{NVIDIA Tesla K40 GPU}.
The portions of the code that ran on \texttt{CPU} were serial, whereas the \texttt{GPU} performed parallel computations.
Protein structures for \texttt{AMOEBA} were prepared using \texttt{pdbxyz} from the \texttt{Tinker} package, and parameterized with \texttt{amoebapro04}\cite{PonderCase2003} or \texttt{amoebapro13},\cite{ShiETal2013} whereas for point-charge force fields, we used \texttt{pdb2pqr}.\cite{Dolinsky04}
From the molecular structure and atomic radii, we generated meshes with the \texttt{msms} software.\cite{MSMS}

\section*{\sffamily \Large Verification and validation}
\subsection*{\sffamily \large Spherical cavity}
Kong and Ponder\cite{KongPonder1997} derived a closed expression for spherical cavities with a random distribution of multipoles, equivalent to the classic Kirkwood solution\cite{Kirkwood1934} for point charges, and suggested using self-consistent iterations to account for polarizability.
We implemented the latter method to verify our numerical implementation in \texttt{PyGBe}.
The test case was a spherical cavity of $R=4$\AA, with a $1$\AA-thick Stern layer, dielectric constant $\epsilon=4$, and three polarizable multipoles with unit charge, dipole, quadrupole, and polarizability, immersed in water ($\epsilon=80$) with salt ($\kappa=0.125$\AA$^{-1}$).
One of the multipoles was placed in the center of the sphere and the other two were off-centered by $2$\AA, as sketched in Figure \ref{fig:sphere_case}.
The solvation energy of such system, computed with Kong and Ponder's expression, is $-65.1131$kcal/mol.
Figure \ref{fig:convergence_sphere} shows the discretization error of the boundary element solution for different mesh densities, which is, as expected, decaying with the average area of the boundary elements.
Both of the surfaces in Figure \ref{fig:sphere_case} have the same amount of mesh elements for this simulation, and the $x$ axis in Figure \ref{fig:convergence_sphere} corresponds to the total number of them.
This result verifies the implementation of \texttt{PyGBe}'s extension to work with polarizable multipoles.

\subsection*{\sffamily \large Comparison with Tinker-APBS}
The work by Schnieders and co-workers \cite{SchniedersBakerRenPonder2007} validates the implementation of polarizable multipoles in \texttt{APBS} using \texttt{1CRN},\cite{Teeter1984} \texttt{1ENH},\cite{ClarkeETal1994} \texttt{1FSV},\cite{DahiyatMayo1997} \texttt{1PGB},\cite{GallagherETal1994} and \texttt{1VII},\cite{McknightETal1997} and comparing their implicit-solvent results with explicit solvent calculations.
Figure \ref{fig:validation_apbs} shows the solvation energy for the same proteins, computed with \texttt{APBS} amd \texttt{PyGBe}, for various grids --- in particular, we used a spacing of $\Delta x=0.61, 0.31, 0.18,$ and $0.16$ for \texttt{APBS} and a mesh with $2, 4, 8$, and $16$ vertices per square angstrom in \texttt{PyGBe}.
\texttt{APBS} is a volumetric solver, whereas \texttt{PyGBe}'s mesh runs only on the molecular surface, hence, mesh sizes are not comparable.
To overcome this situation, we performed Richardson extrapolation\cite{Roache1998,CooperBardhanBarba2014} and obtained an approximation of the exact solution from the numerical calculations, as the mesh density tends to infinity. 
For \texttt{APBS} runs, we used meshes with $\Delta x=0.61, 0.31,$ and $0.16$ to compute the extrapolated values, whereas in the case of \texttt{PyGBe}, we used grids with $2$, $4$, and $8$ vertices per square angstrom.
Table \ref{table:extrapolation} presents the extrapolated values for each case, and they are also marked with a dotted line in Figure \ref{fig:validation_apbs}. 
Using the extrapolated values as an \emph{exact} solution, we computed the error in solvation energy for each mesh, and plotted them against the time to solution in Figure \ref{fig:time_convergence}. 
Moreover, from Richardson extrapolation we are also able to compute an \emph{observed order of convergence}, which is the rate of convergence of simulations towards the extrapolated value, and they are also included in Table \ref{table:extrapolation}.

We further compared \texttt{APBS} and \texttt{PyGBe} using the total dipole moment of the protein. 
In our simulations, we saw that the mesh density had a very weak effect on the total dipole moment, hence, Table \ref{table:dipole} only shows results for the coarsest meshes in each case ($\Delta x = 0.61$ in \texttt{APBS} and 2 vertices per square angstrom in \texttt{PyGBe}).

For consistency with Schnieders' work,\cite{SchniedersBakerRenPonder2007} these runs were performed using the \texttt{amoebapro04} force field, a protein dielectric region of $\epsilon_\text{prot}=1$ and a solvent with dielectric constant $\epsilon_\text{solv}=78.3$ and $150$mM of salt ($\kappa = 0.125$\AA$^{-1}$).
The tolerance of the self-consistent solver for the induced dipole moment was set to $10^{-2}$ and the exit criterion of the \texttt{GMRES} in the Poisson-Boltzmann linear solver to $10^{-5}$.

In the \texttt{APBS} simulations, we used a sharp surface definition (keyword \texttt{SMOL}) rather than the smoothed definition based on fourth order splines used by Schnieders and co-workers\cite{SchniedersBakerRenPonder2007} (keyword \texttt{SPL4}), which explains the differences in the solvation energy between the latter work and Figure \ref{fig:validation_apbs}.
The \texttt{SMOL} definition is closer to the surface description from a boundary integral formulation.

For the \texttt{PyGBe} simulations, we used 1 Gauss quadrature point per boundary element far-away from the collocation point, however, for nearly-singular integrals (within $1.25\sqrt{A}$\AA~ of the collocation point) we used finer quadrature rules with 19 Gauss nodes. 
With respect to the treecode acceleration, we set the multipole acceptance criterion to $\theta=0.6$, and used Taylor expansions up to order $P=4$. 
For efficiency, the tree was built making sure that no box of the lowest level had more than $50$ elements for \texttt{CPU} and $300$ elements for \texttt{GPU}.

\section*{\sffamily \Large Influence of size}

From Figure \ref{fig:time_convergence}, we can see that a mesh density with 4 vertices per square angstroms yields a solution that is around 1\% away from a converged value.
To study the behavior of \texttt{PyGBe} with respect to the size of the protein, we computed the solvation energy of \texttt{1PGB},\cite{GallagherETal1994} \texttt{1LYZ},\cite{Diamond1974} \texttt{1A7M},\cite{HindsETal1998} \texttt{1X1U},\cite{IkuraETal2004} and \texttt{1IGT}\cite{HarrisETal1997} using 4 vertices per square angstrom, and report them in Table \ref{table:size}.
These runs were done using the \texttt{amoebapro13}\cite{ShiETal2013} force field, and with the same parameters of the \texttt{GPU} runs that led to Figure \ref{fig:validation_apbs}.

\section*{\sffamily \Large Comparison with standard force fields}
The following simulations compare \texttt{AMOEBA} (with \texttt{amoebapro13}) with standard point-charge force fields, such as \texttt{AMBER}, \texttt{CHARMM}, and \texttt{PARSE}.

\subsection*{\sffamily \large Mesh refinement study}
We performed a further mesh convergence study of the solvation energy of protein GB1 (\texttt{1PGB}) using \texttt{AMOEBA}, and the fixed-charge force field \texttt{AMBER}, with the same simulation parameters that led to the results in Figure \ref{fig:validation_apbs}.
Figure \ref{fig:pgb_comparison} shows the solvation energy with both force fields decaying approximately with the average area, which is the expected behavior for a piecewise constant boundary element solution.\cite{CooperBardhanBarba2014}
In particular, the observed order con convergence for \texttt{AMOEBA} is $\mathcal{O}(A^{1.22})$ and for \texttt{AMBER} it is $\mathcal{O}(A^{0.75})$, whereas the Richardson extrapolated values are $-812.06$ kcal/mol and $-959.07$ kcal/mol, respectively.
Also, Figure \ref{fig:pgb_comparison} shows the time to solution for different meshes using \texttt{AMOEBA} and \texttt{AMBER}, with the expected $\mathcal{O}(N\log N)$ scaling from the treecode, which is the dominant part of the algorithm.

\subsection*{\sffamily \large Binding energy calculations}
We computed the electrostatic component of the absolute and relative binding energies for the \texttt{HIV-1 GP120} core complexed with \texttt{CD4} and a neutralizing human antibody (\texttt{PDB: 1GC1}\cite{KwongETal1998}).
\texttt{CD4} is a glycoprotein present on the surface of immune cells, and induces a binding site for the monoclonal antibody on \texttt{GP120}.
Here, we compute the binding energy of \texttt{GP120} (that has \texttt{CD4} already attached to it, totalling $7425$ atoms) with the antigen-binding fragment (\texttt{Fab}) of the antibody ($6667$ atoms), which generates a complex with $14092$ atoms.

Table \ref{table:bind} shows the results for binding energy calculations computed with \texttt{AMOEBA} (with and without polarizability), \texttt{AMBER}, \texttt{CHARMM}, and \texttt{PARSE}, using Equation \eqref{eq:bind_energy} and Equation \eqref{eq:rel_bind_energy}.
The mesh density for these runs was $4$ vertices per square angstrom, and we used the same simulation parameters that led to the results in Figure \ref{fig:validation_apbs}, only differing by setting the multipole acceptance criterion to $\theta=0.5$, and the threshold between near and far-away integrals at $2\sqrt{A}$.
For the case of point-charge force fields, we varied the dielectric constant of the protein ($\epsilon_\text{prot}$) from $1$ to $6$, to test if it was possible to implicitly capture polarization in those cases.

\section*{\sffamily \Large DISCUSSION}
\section*{\sffamily \Large Verification and validation}
The verification result for the sphere in Figure \ref{fig:convergence_sphere} shows the numerical value from \texttt{PyGBe} approaching the analytical solution from Kong and Ponder\cite{KongPonder1997} with the average area of the surface mesh elements, which is the expected behavior for a piecewise constant boundary element approximation.\cite{CooperBardhanBarba2014} 
This indicates that the implementation of polarizable multipoles on \texttt{PyGBe} is correctly solving the mathematical model.
With that, we further validate \texttt{PyGBe} by comparing our numerical results with the implementation from Schnieders and co-workers\cite{SchniedersBakerRenPonder2007} in Figure \ref{fig:validation_apbs} and Table \ref{table:extrapolation}.
For well converged simulations, the observed order of convergence is expected to match the order of the method: $\mathcal{O}(\Delta x)$ for \texttt{APBS} and $\mathcal{O}(\text{area})$ for \texttt{PyGBe},\cite{CooperBardhanBarba2014} however, this is not the case in Table \ref{table:extrapolation}, specially for \texttt{APBS}.  
This is probably due to the difficult transfer of high order multipole charges to the finite-difference mesh.
In \texttt{PyGBe}, slight deviations to the expected slope are forseeable due to the irregular nature of the surface mesh generated by \texttt{msms}, as it is impossible to refine it homogeneously. 
Regardless of these shortcomings, the extrapolated values of solvation energy for \texttt{APBS} and \texttt{PyGBe} are in good agreement (less than 1\% off), proving that the main features of the electrostatic potential are correctly represented in both cases.

Figure \ref{fig:time_convergence} shows how time scales with the error in the simulation.
The expected behavior is faster simulations with \texttt{APBS} at low accuracy that scale worse than \texttt{PyGBe} as the mesh refines,\cite{CooperBardhanBarba2014} yet, both codes present similar slopes. 
This unexpected trend is an artifact of having a high observed order of convergence with \texttt{APBS}, but the plots are still useful to perform a fair comparison of the volumetric and boundary integral solvers.
All \texttt{CPU} runs are single-core, hence, \texttt{APBS} and \texttt{PyGBe-CPU} timings in Figure \ref{fig:time_convergence} are comparable, yet, it is unfair to compare those with \texttt{GPU} run times.
From these results, we can conclude that \texttt{PyGBe} and \texttt{AMOEBA} have equivalent performance under the same \texttt{CPU} conditions. 

We still plotted \texttt{GPU} timings for reference, considering that the extension of \texttt{PyGBe} to use \texttt{AMOEBA} is implemented in both \texttt{CPU} and \texttt{GPU}.
From here, we can see an advantage of the boundary integral approach, as it performs well on graphics cards.

\section*{\sffamily \Large Influence of size}
Table \ref{table:size} shows that the size of the biomolecule has a weak effect on the number of self-consistent iterations required for the induced dipoles to converge. 
Between \texttt{1PGB} and \texttt{1IGT} the number of atoms increases by $20\times$, however, only one extra iteraton was required. 
This suggests that this implicit-solvent Poisson-Boltzmann approach is efficient to analyze large biomolecular systems.
Moreover, the solvent-excluded surface mesh grows slower than a volumetric mesh as the molecule size increases, indicating that a boundary integral approach is more appropriate for a large-scale application.

\section*{\sffamily \Large Comparison with standard force fields}

\subsection*{\sffamily \large Mesh refinement study}
Figure \ref{fig:pgb_comparison} is a further mesh convergence study of \texttt{1PGB}, comparing \texttt{AMOEBA} with \texttt{AMBER}.
Results with both force fields are converging as expected (observed order of convergence close to $\mathcal{O}(A)$), proving that the electrostatic potential is correctly resolved regardless of the charge description.
This means that the dipole and quadrupole components in \texttt{AMOEBA} do not affect the mesh density required to resolve the potential field appropriately, and the same mesh sizes can be used in both cases.

The Richardson extrapolated value of solvation energy with \texttt{AMBER} is $-959.07$ kcal/mol and with \texttt{AMOEBA} is $812.06$ kcal/mol, differring by around $15$\%. 
There are a number of reasons that could explain this difference, for example, the fact that the parameterization of the protein might not be optimal.
Also, in point-charge force fields, slightly higher ad-hoc permittivities are usually used inside the protein\cite{GilsonHonig1986} to account for reorentation of dipoles, and the solvation energy with \texttt{AMBER} decreases with a higher dielectric constant, approaching the value for \texttt{AMOEBA}.

\subsection*{\sffamily \large Binding energy calculations}
Even though the relative differences in solvation energy may not be very large, they can be critical when studying changes of this quantity, for example, in binding energy calculations.
We find evidence of this in Table \ref{table:bind}, where the solvation energies of the complex with \texttt{AMBER} and \texttt{AMOEBA} differ by less than $2\%$, however, the absolute binding energy is off by $25\%$.

Table \ref{table:bind} shows smaller binding energies with \texttt{AMOEBA}, compared to its point-charge counterparts using $\epsilon_\text{prot}=1$.
This behaviour is expected.
In fact, in point-charge force fields polarization is implicitly considered with higher permittivities, which, as shown in Table \ref{table:bind}, results in a lower absolute binding energy.
This effect is already considered in a polarizable force field with a relative permittivity inside the protein of $\epsilon_\text{prot}=1$.

Regardless of these differences, Table \ref{table:bind} consistently shows that the electrostatic component of the absolute binding energy in dissolved state ($\Delta G_\text{bind,diss}$) is positive, hence, repulsive.
Furthermore, the values of binding energy are similar, which indicates that the parameterization with all force fields is adequate.
From here, we can conclude that the nonpolar component of the solvation energy is key to correctly predict the expected binding of the antibody to the \texttt{CD4}-induced binding cite of \texttt{GP120}.\cite{KwongETal1998}

Even though all force fields arrive to similar values of binding energy, it is interesting to analyze each component of the thermodynamic cycle in Figure \ref{fig:thermo_cycle} separately.
For the case of \texttt{AMOEBA}, we see a negative relative binding energy ($\Delta\Delta G_\text{bind}<0$), which means that binding is more likely to happen in dissolved state than vacuum (see Equation \eqref{eq:rel_bind_energy}).
Hence, the role of the solvent is to enhance attraction, and the repulsion is given by the energetic difference in vacuum.
On the other hand, all point charge force fields predict the opposite physical phenomena: the attraction of the biomolecules is due to coulombic-type interactions ($\Delta G_\text{bind,vac}<0$), and the solvent induces repulsion ($\Delta\Delta G_\text{bind}>0$).
Hence, though we are predicting correct absolute binding energies, we may be misled in what is the role of each term of the thermodynamic cycle.

Also, in Table \ref{table:bind}, the relative binding energy of \texttt{AMOEBA} drops in half when polarizability is not considered.
This shows that an important part of the binding energy comes from polarization, and it is critical to consider it in applications such as this one, where the two biomolecules polarize each other.

\section*{\sffamily \Large CONCLUSIONS}

This paper presents an extension of the Poisson-Boltzmann solver \texttt{PyGBe} to use a charge distribution with high order multipoles and polarizability, in particular, through the \texttt{AMOEBA} force field.
The software is based on a boundary integral representation of the partial differential equations, where the molecular charge is considered on the right-hand side of the resulting linear system, making it relatively easy to implement in an existing code base.
We verified this extension against closed expressions valid for spherical inclusions, and validated it by contrasting with a similar implementation that uses \texttt{APBS}.
\texttt{PyGBe} and \texttt{APBS} perform equivalently in serial \texttt{CPU} runs, however, the boundary integral approach presents an important speedup on the \texttt{GPU}. 
We also compared the performance of \texttt{AMOEBA} and standard point-charge force fields, like \texttt{AMBER}, in a Poisson-Boltzmann model. 
In that comparison, we realized that the same mesh densities can appropriately resolve the electrostatic potential in either case, and that they converge to similar results.
Also, considering polarizability can be extremely important in situations that have cooperative effects, such as binding, where the two molecules are mutually polarized.
In that case, we saw that even though an appropriate parameterization of a force field may yield the correct values of the absolute binding energy, there are differences in the mechanisms present in the interaction, where for point-charge force fields the solvent induces repulsion, whereas for \texttt{AMOEBA} it favors binding. 

We conclude that this boundary-integral implicit-solvent approach can efficiently compute the electrostatic potential in biomolecular systems with polarizable force fields. 
The fact that the charge distribution is computed analytically on the right-hand side avoids any difficult point-multipole transfer to the mesh, making a boundary-integral representation ideal for these computations, specially looking at large-scale simulations.


\subsection*{\sffamily \large ACKNOWLEDGMENTS}

The author is grateful of the helpful interactions with Prof. Robert Krasny (UMich), Prof. Weihua Geng (SMU), and Prof. Michael Schnieders (UIowa).
Part of this work was done while visiting the University of Michigan, hosted by Prof. Robert Krasny.
The research was financially supported by CONICYT though FONDECYT Iniciaci\'on N$^\circ$ 11160768, and Basal Project FB 0821.


\section*{\sffamily \Large APPENDIX}
This appendix shows the detailed expressions of the derivatives needed in different calculations.

The first derivatives of the Green's function required in Equation \eqref{eq:Efield_solv} to obtain the electric field due to the solvent reaction and Equation \eqref{eq:energy_multipole} for solvation energy are
\begin{align} \label{eq:Efield_solv_der}
\frac{\partial G_L}{\partial r_i} (\mathbf{r},\mathbf{r}') &= \frac{\partial}{\partial r_i} \left(\frac{1}{|\mathbf{r}-\mathbf{r}'|}\right) = -\frac{r_i-r'_i}{|\mathbf{r}-\mathbf{r}'|^3} \nonumber \\
\frac{\partial}{\partial r_i}\frac{\partial G_L}{\partial \mathbf{n}'} (\mathbf{r},\mathbf{r}') &=  \frac{\partial}{\partial r_i} \frac{\partial}{\partial r'_j} \left(\frac{1}{|\mathbf{r}-\mathbf{r}'|}\right)n'_j \nonumber \\
& = \left(\frac{\delta_{ij}}{|\mathbf{r}-\mathbf{r}'|^3} - 3\frac{(r_j-r'_j)(r_i-r'_i)}{|\mathbf{r}-\mathbf{r}'|^5}\right)n'_j,
\end{align}
where the normal $\mathbf{n}'=(n'_1,n'_2,n'_3)$ moves with the integration variable $\mathbf{r}'$.
On the other hand, the second derivatives used in Equation \eqref{eq:ddphi}  are
\begin{align}\label{eq:sec_derivatives}
&\frac{\partial}{\partial r_k}\frac{\partial}{\partial r_i}G_L(\mathbf{r},\mathbf{r}') = \frac{\partial}{\partial r_k}\frac{\partial}{\partial r_i}\left( \frac{1}{|\mathbf{r}-\mathbf{r}'|}\right) = \left( -\frac{\delta_{ij}}{|\mathbf{r}-\mathbf{r}'|^3} + 3\frac{(r_k-r'_k)(r_i-r'_i)}{|\mathbf{r}-\mathbf{r}'|^5} \right) \nonumber \\
&\frac{\partial}{\partial r_k}\frac{\partial}{\partial r_i} \frac{\partial G_L}{\partial \mathbf{n}'}(\mathbf{r},\mathbf{r}') = \frac{\partial}{\partial r_k}\frac{\partial}{\partial r_i} \frac{\partial}{\partial r'_j} \left( \frac{1}{|\mathbf{r}-\mathbf{r}'|}\right) n'_j \nonumber \\
&= -\frac{3}{|\mathbf{r}-\mathbf{r}'|^5}\left[(r_k-r'_k)\delta_{ij} + (r_i-r'_i) \delta_{jk} +(r_j-r'_j)\delta_{ik} -5\frac{(r_i-r'_i)(r_j-r'_j)(r_k-r'_k)}{|\mathbf{r}-\mathbf{r}'|^2} \right] n'_j.
\end{align}

The derivatives of the terms in Equation \eqref{eq:Efield_mult} are:
\begin{align} \label{eq:derivatives_Emult}
\frac{\partial}{\partial r^{(l)}_k} \left(\frac{1}{|\mathbf{r}^{(l)}-\mathbf{r}^{(m)}|} \right) &= -\frac{r_k^{(l)}-r_k^{(m)}}{|\mathbf{r}^{(l)}-\mathbf{r}^{(m)}|^3}, \nonumber \\
\frac{\partial}{\partial r^{(l)}_k} \left( \frac{r^{(l)}_i-r^{(m)}_i}{|\mathbf{r}^{(l)}-\mathbf{r}^{(m)}|^3} \right) &= \frac{\delta_{ik}}{|\mathbf{r}^{(l)}-\mathbf{r}^{(m)}|^3} - 3\frac{(r_i^{(l)}-r_i^{(m)})(r_k^{(l)}-r_k^{(m)})}{|\mathbf{r}^{(l)}-\mathbf{r}^{(m)}|^5} \text{, and} \nonumber \\
\frac{\partial}{\partial r^{(l)}_k} \left( \frac{(r_i^{(l)}-r_{i}^{(m)})(r_{j}^{(l)}-r_{j}^{(m)})}{|\mathbf{r}^{(l)}-\mathbf{r}^{(m)}|^5} \right) &= -5 \frac{(r_i^{(l)}-r_i^{(m)})(r_j^{(l)}-r_j^{(m)})(r_k^{(l)}-r_k^{(m)})}{|\mathbf{r}
^{(l)}-\mathbf{r}^{(m)}|^7} \nonumber \\ 
&+ \frac{(r_i^{(l)}-r_i^{(m)})\delta_{jk}}{|\mathbf{r}^{(l)}-\mathbf{r}^{(m)}|^5} + \frac{(r_j^{(l)}-r_j^{(m)})\delta_{ik}}{|\mathbf{r}^{(l)}-\mathbf{r}^{(m)}|^5},
\end{align}

The terms required to compute the second derivative of the potential due to a collection of multipoles are
\begin{align}\label{eq:second_derivative}
\frac{\partial}{\partial r^{(l)}_i} \frac{\partial}{\partial r^{(l)}_j} \left( \frac{1}{|\mathbf{r}^{(l)}-\mathbf{r}^{(m)}|} \right) &= -\frac{\delta_{ij}}{|\mathbf{r}^{(l)}-\mathbf{r}^{(m)}|^3} + 3\frac{(r_i^{(l)}-r_i^{(m)})(r_j^{(l)}-r_j^{(m)})}{|\mathbf{r}^{(l)}-\mathbf{r}^{(m)}|^5} \nonumber \\
\frac{\partial}{\partial r^{(l)}_k}\frac{\partial}{\partial r^{(l)}_j} \left( \frac{r^{(l)}_i-r^{(m)}_i}{|\mathbf{r}^{(l)}-\mathbf{r}^{(m)}|^3} \right) &= \frac{3}{|\mathbf{r}^{(l)}-\mathbf{r}^{(m)}|^5} \left( 5\frac{(r_i^{(l)}-r_i^{(m)})(r_j^{(l)}-r_j^{(m)})(r_k^{(l)}-r_k^{(m)})}{|\mathbf{r}^{(l)}-\mathbf{r}^{(m)}|^2}\right. \nonumber\\
&\left.- (r_j^{(l)}-r_j^{(m)})\delta_{ik} - (r_i^{(l)}-r_i^{(m)})\delta_{jk} - (r_k^{(l)}-r_k^{(m)})\delta_{ij}\right) \nonumber\\
\frac{\partial}{\partial r^{(l)}_o}\frac{\partial}{\partial r^{(l)}_k} \left( \frac{(r_i^{(l)}-r_{i}^{(m)})(r_{j}^{(l)}-r_{j}^{(m)})}{|\mathbf{r}^{(l)}-\mathbf{r}^{(m)}|^5} \right) &= -\frac{5}{|\mathbf{r}^{(l)}-\mathbf{r}^{(m)}|^7}\nonumber\\
&\left[ -7\frac{(r_i^{(l)}-r_i^{(m)})(r_j^{(l)}-r_j^{(m)})(r_k^{(l)}-r_k^{(m)})(r_o^{(l)}-r_o^{(m)})}{|\mathbf{r}^{(l)}-\mathbf{r}^{(m)}|^2}\right.\nonumber\\
&\left. +(r_j^{(l)}-r_j^{(m)})(r_o^{(l)}-r_o^{(m)})\delta_{ik} + (r_k^{(l)}-r_k^{(m)})(r_j^{(l)}-r_j^{(m)})\delta_{oi} \right.\nonumber\\ 
&+\left.(r_i^{(l)}-r_i^{(m)})(r_k^{(l)}-r_k^{(m)})\delta_{jo} + (r_o^{(l)}-r_o^{(m)})(r_i^{(l)}-r_i^{(m)})\delta_{kj}\right.\nonumber\\
&\left. + (r_j^{(l)}-r_j^{(m)})(r_i^{(l)}-r_i^{(m)})\delta_{ok}\right]
\end{align}
%


\clearpage


\bibliography{compbio,fastmethods,scicomp,vortexmeth,scbib,cfd,bem}


\clearpage

\begin{figure}
\caption{\label{cc} Place Figure 1 caption here. In the case of reproduced figures in review articles, you must obtain the publisher's permission and state a suitable notice here along with a citation.}
\end{figure}

\begin{figure}
\caption{\label{fig:biomolecule} Graphical representation of the implicit-solvent model.}
\end{figure}

\begin{figure}
\caption{\label{fig:thermo_cycle} Thermodynamic cycle for the calculation of binding energy.}
\end{figure}

\begin{figure}
\caption{\label{fig:sphere_case} Graphical representation of the spherical model used for validation.}
\end{figure}

\begin{figure}
\caption{\label{fig:convergence_sphere} Mesh refinement study for the system in Figure \ref{fig:sphere_case}.}
\end{figure}

\begin{figure}
\caption{\label{fig:validation_apbs} Solvation energy for different mesh sizes. The dotted line represents the Richardson extrapolation from Table \ref{table:extrapolation}.}
\end{figure}

\begin{figure}
\caption{\label{fig:time_convergence} Time to solution versus errors for different mesh sizes. Errors were computed using the values of Table \ref{table:extrapolation} as the reference.}
\end{figure}

\begin{figure}
\caption{\label{fig:pgb_comparison} Convergence and timing comparison between \texttt{AMBER} and \texttt{AMOEBA}}
\end{figure}

%


\clearpage

\begin{center}
\includegraphics[width=0.2\columnwidth,keepaspectratio=true]{figures/intro_plot.pdf}
\end{center}
\vspace{0.25in}
\hspace*{3in}
{\Large
\begin{minipage}[t]{3in}
\baselineskip = .5\baselineskip
Figure 1 \\
Christopher D. Cooper \\
J.\ Comput.\ Chem.
\end{minipage}
}
\clearpage

\begin{center}
\includegraphics[width=0.4\textwidth]{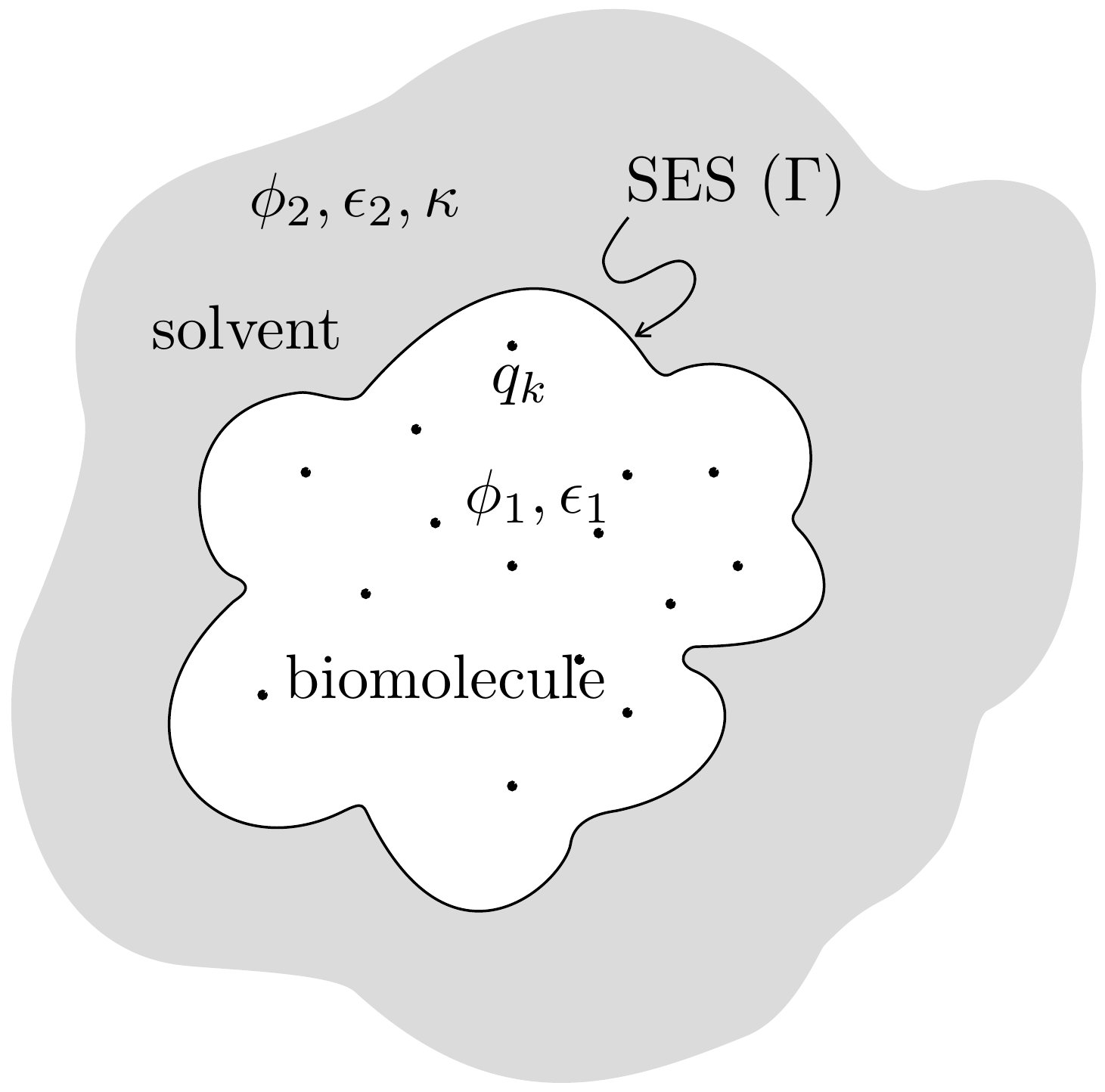}
\end{center}
\vspace{0.25in}
\hspace*{3in}
{\Large
\begin{minipage}[t]{3in}
\baselineskip = .5\baselineskip
Figure 2 \\
Christopher D. Cooper \\
J.\ Comput.\ Chem.
\end{minipage}
}

\clearpage

\begin{center}
\includegraphics[width=0.5\columnwidth,keepaspectratio=true]{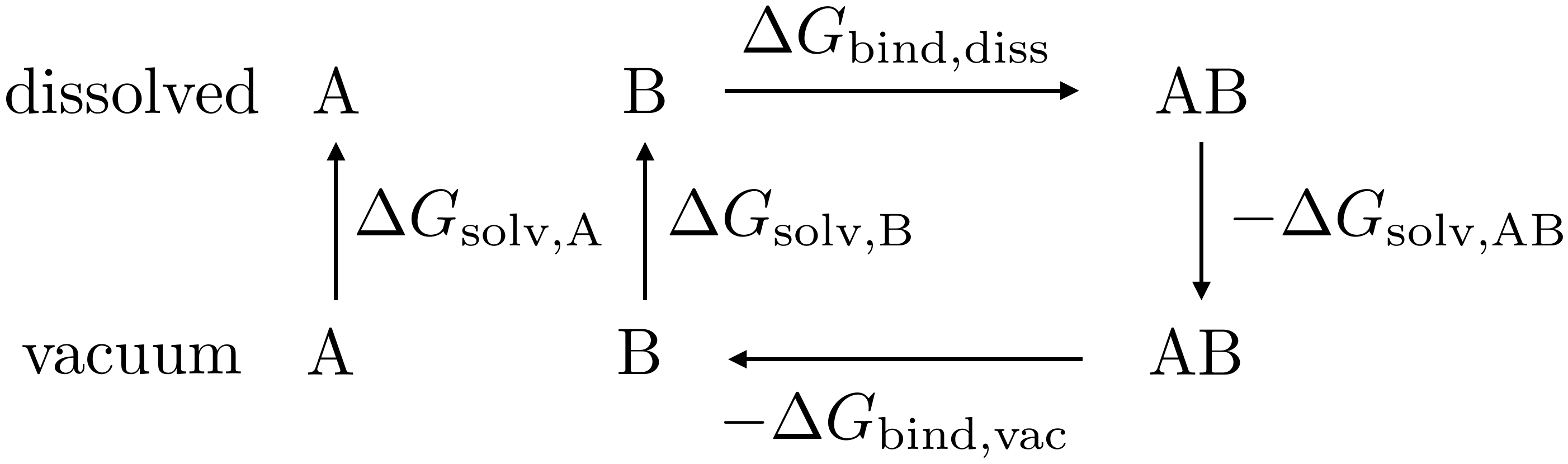}
\end{center}
\vspace{0.25in}
\hspace*{3in}
{\Large
\begin{minipage}[t]{3in}
\baselineskip = .5\baselineskip
Figure 3 \\
Christopher D. Cooper \\
J.\ Comput.\ Chem.
\end{minipage}
}
\clearpage

\begin{center}
\includegraphics[width=\textwidth]{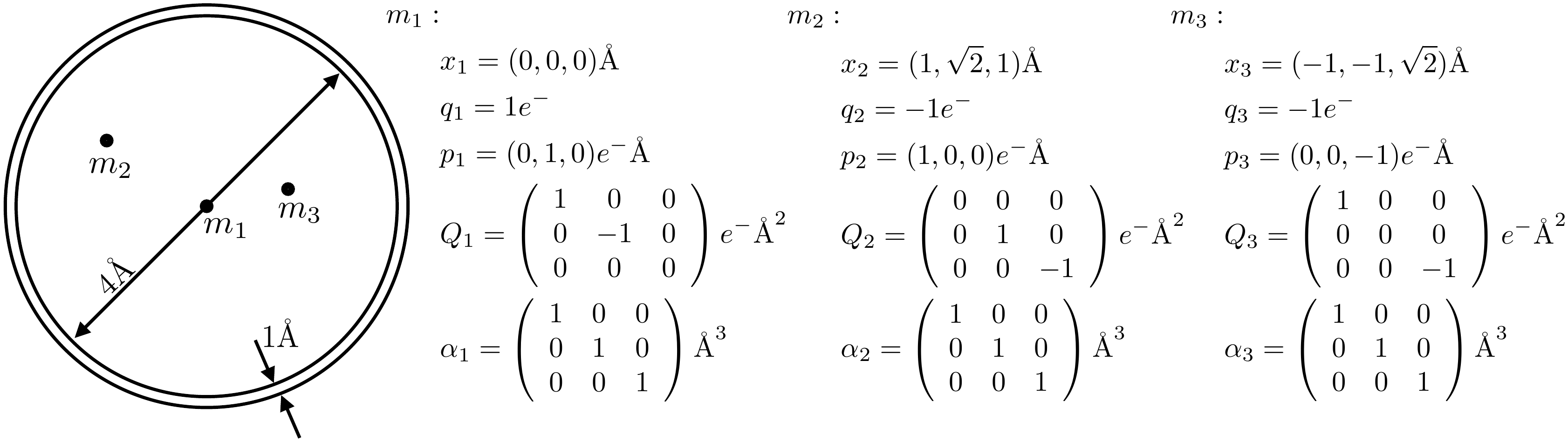}
\end{center}
\vspace{0.25in}
\hspace*{3in}
{\Large
\begin{minipage}[t]{3in}
\baselineskip = .5\baselineskip
Figure 4 \\
Christopher D. Cooper \\
J.\ Comput.\ Chem.
\end{minipage}
}

\clearpage

\begin{center}
\includegraphics[width=0.5\textwidth]{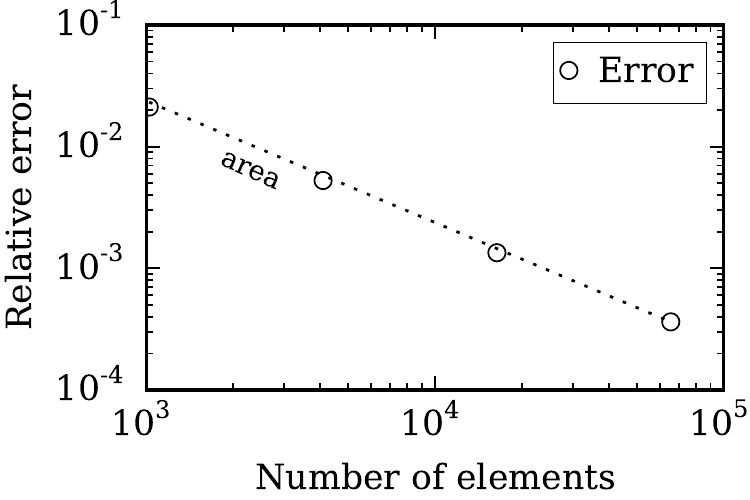}
\end{center}
\vspace{0.25in}
\hspace*{3in}
{\Large
\begin{minipage}[t]{3in}
\baselineskip = .5\baselineskip
Figure 5 \\
Christopher D. Cooper \\
J.\ Comput.\ Chem.
\end{minipage}
}

\clearpage

\begin{center}
\includegraphics[width=0.4\textwidth]{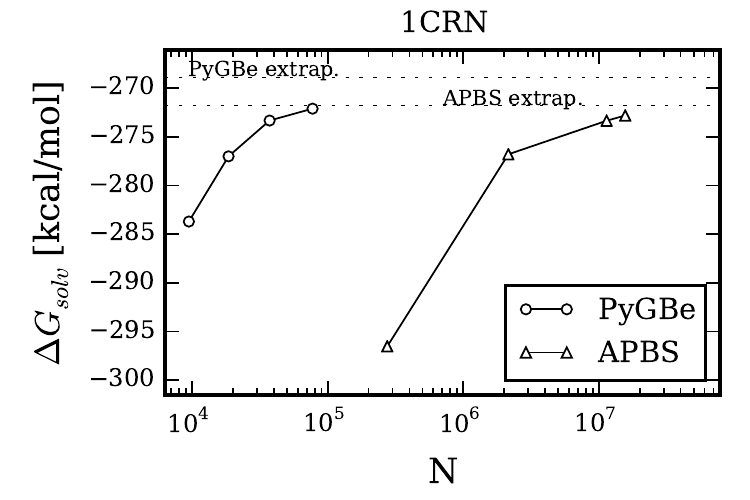}
\includegraphics[width=0.4\textwidth]{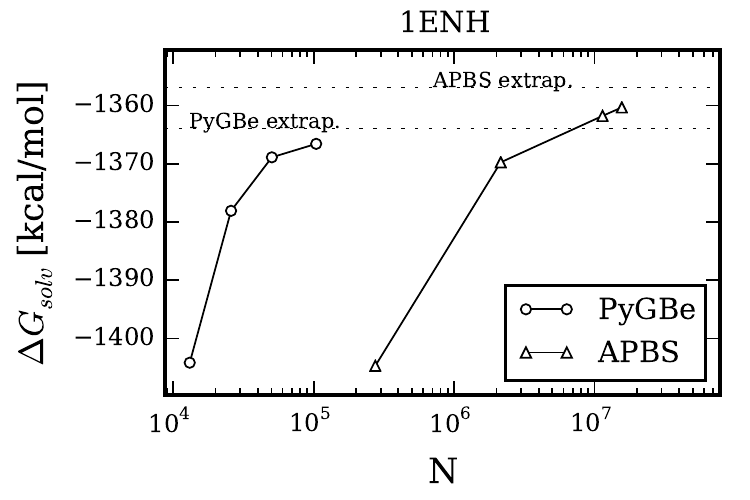}
\includegraphics[width=0.4\textwidth]{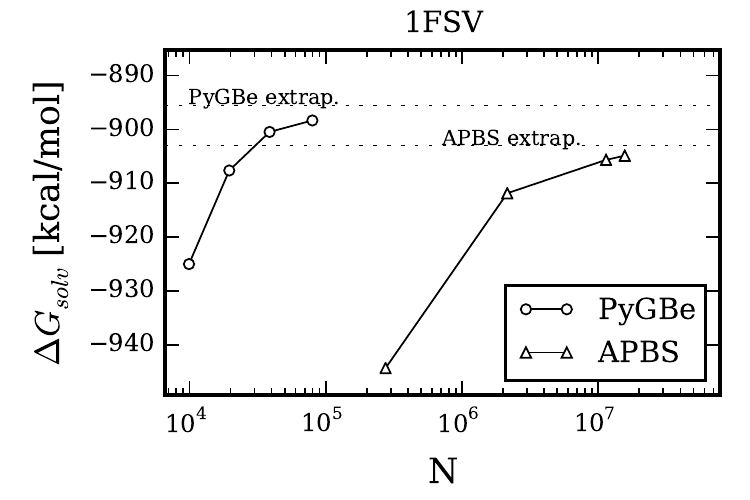}
\includegraphics[width=0.4\textwidth]{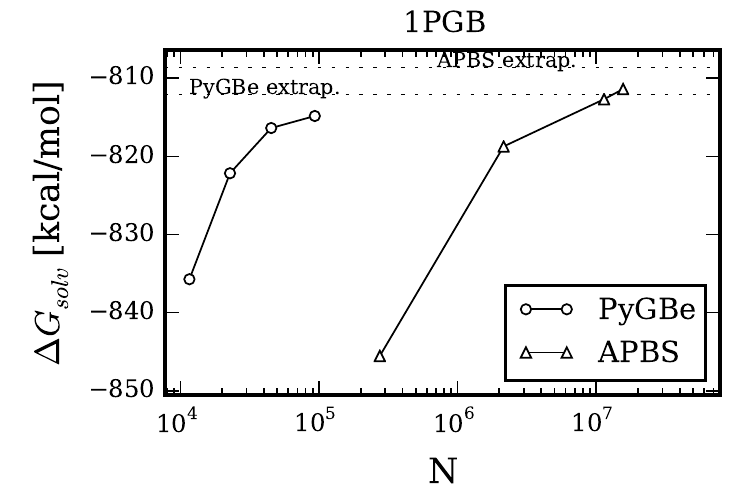}
\includegraphics[width=0.4\textwidth]{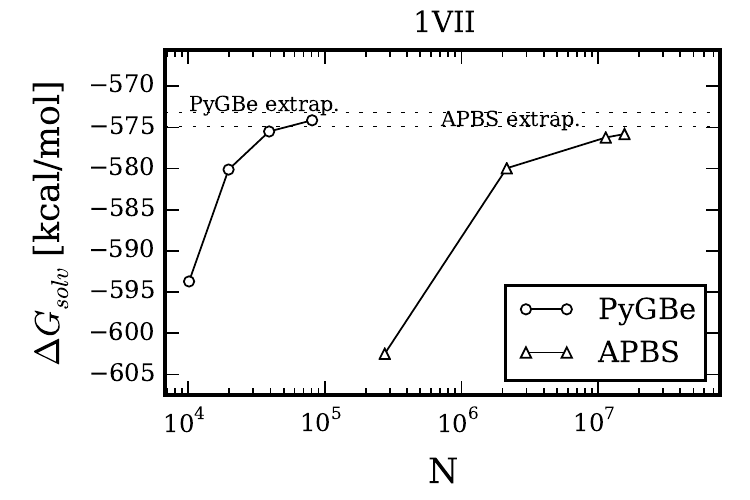}
\end{center}
\vspace{0.25in}
\hspace*{3in}
{\Large
\begin{minipage}[t]{3in}
\baselineskip = .5\baselineskip
Figure 6 \\
Christopher D. Cooper \\
J.\ Comput.\ Chem.
\end{minipage}
}

\clearpage

\begin{center}
\includegraphics[width=0.4\textwidth]{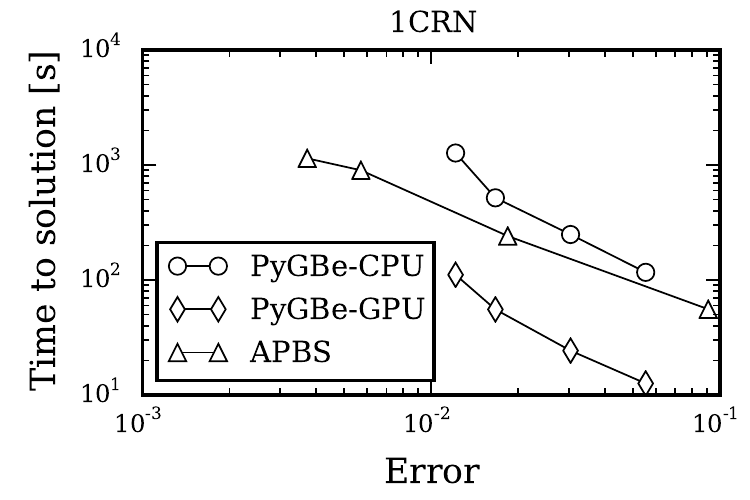}
\includegraphics[width=0.4\textwidth]{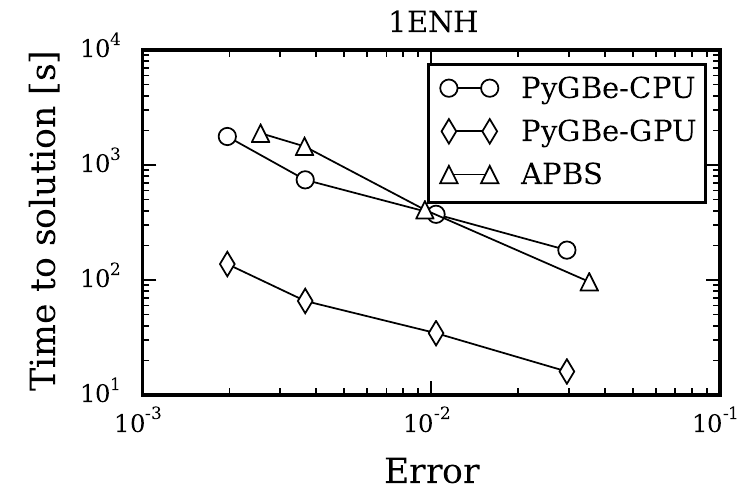}
\includegraphics[width=0.4\textwidth]{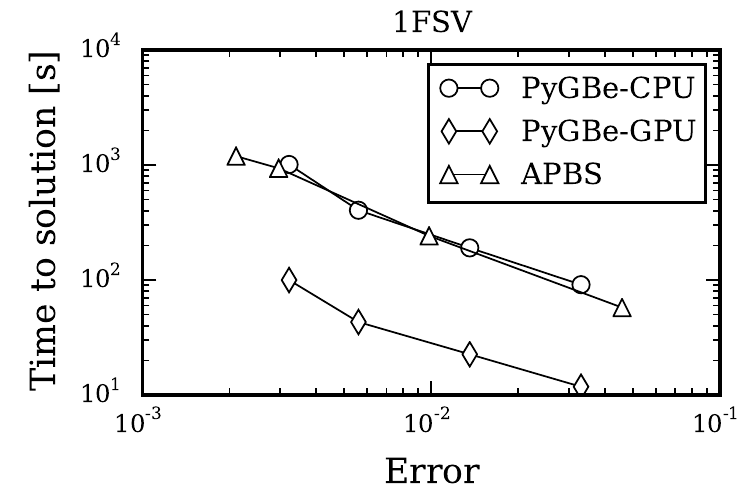}
\includegraphics[width=0.4\textwidth]{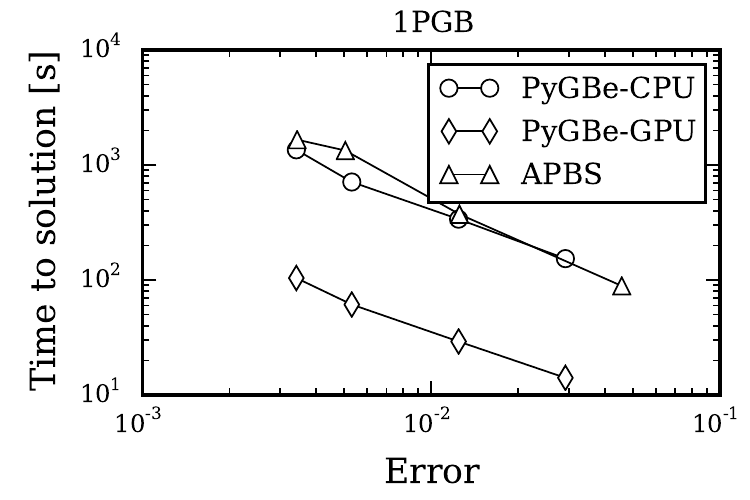}
\includegraphics[width=0.4\textwidth]{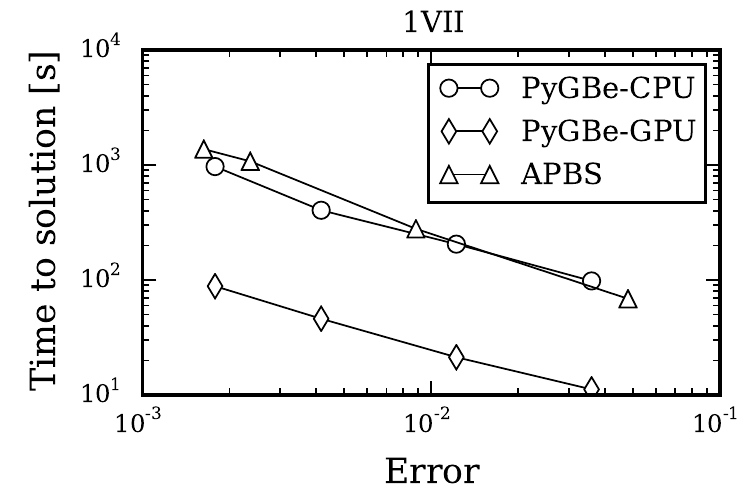}
\end{center}
\vspace{0.25in}
\hspace*{3in}
{\Large
\begin{minipage}[t]{3in}
\baselineskip = .5\baselineskip
Figure 7 \\
Christopher D. Cooper \\
J.\ Comput.\ Chem.
\end{minipage}
}
\clearpage

\begin{center}
\includegraphics[width=0.4\textwidth]{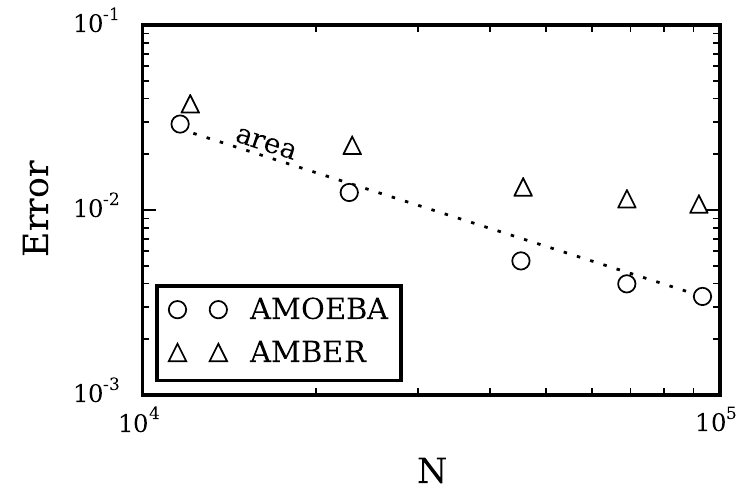}
\includegraphics[width=0.4\textwidth]{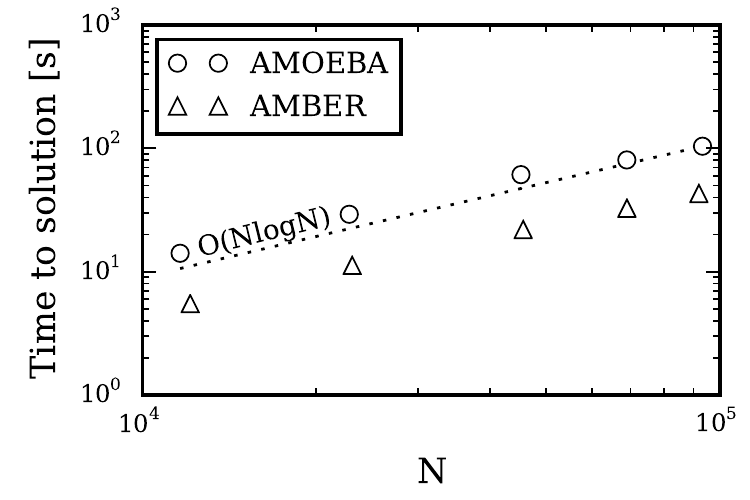}
\end{center}
\vspace{0.25in}
\hspace*{3in}
{\Large
\begin{minipage}[t]{3in}
\baselineskip = .5\baselineskip
Figure 8 \\
Christopher D. Cooper \\
J.\ Comput.\ Chem.
\end{minipage}
}
\clearpage

\begin{table}
\begin{tabular}{|c|c c|c c|}\hline
\textbf{Protein} & \multicolumn{2}{c|}{\textbf{$\Delta$G$_\text{solv}$ [kcal/mol]}} & \multicolumn{2}{c|}{\textbf{Convergence}} \\ 
    & \texttt{APBS} & \texttt{PyGBe} & \texttt{APBS} & \texttt{PyGBe} \\ \hline
  1CRN & -271.80 & -268.85 & $\mathcal{O}(\Delta x^{2.30})$ & $\mathcal{O}(A^{1.60})$ \\ \hline
  1ENH & -1356.87 & -1363.9 & $\mathcal{O}(\Delta x^{1.89})$ & $\mathcal{O}(A^{1.50})$ \\ \hline
  1FSV & -902.99 & -895.46 & $\mathcal{O}(\Delta x^{2.22})$ & $\mathcal{O}(A^{1.28})$ \\ \hline
  1PGB & -809.94 & -812.06 & $\mathcal{O}(\Delta x^{2.02})$ & $\mathcal{O}(A^{1.22})$ \\ \hline
  1VII & -574.87 & -573.12 & $\mathcal{O}(\Delta x^{2.44})$ & $\mathcal{O}(A^{1.55})$ \\ \hline
\end{tabular}
\caption{\label{table:extrapolation} Richardson extrapolated solvation energy and observed order of convergence.}
\end{table}

\begin{table}
\begin{tabular}{|c|c c c|}\hline
                  & \multicolumn{3}{c|}{\textbf{Total dipole [Debye]}} \\
\textbf{PDB code} & \textbf{Vacuum} & \texttt{APBS} & \texttt{PyGBe}  \\ \hline 
  1CRN & 62.54 & 82.02 & 82.60  \\ \hline 
  1ENH & 209.94 & 266.25 & 267.02  \\ \hline 
  1FSV & 184.20 & 210.70 & 207.80  \\ \hline 
  1PGB & 101.38 & 130.55 & 129.37  \\ \hline 
  1VII & 158.72 & 194.57 & 194.21 \\ \hline 
\end{tabular}
\caption{\label{table:dipole} Total dipole moment in vacuum and dissolved sate, calculated with \texttt{APBS} and \texttt{PyGBe}.}
\end{table}

\begin{table}
\begin{tabular}{|c|c|c|c|c|c|}\hline
\textbf{PDB code} & \textbf{$N_\text{atoms}$} & \textbf{$N_\text{elements}$} & \textbf{$\Delta$G$_\text{solv}$ [kcal/mol]} & \textbf{Time [s]} & \textbf{$N_\text{iter}$} \\ \hline 
  1PGB & 927 & 24634 & -822.42 & 31.6 & 4 \\ \hline 
  1LYZ & 1961 & 42544 & -1597.47 & 64.9 & 4 \\ \hline 
  1A7M & 2809 & 60582 & -2587.44 & 129.1 & 4 \\ \hline 
  1X1U & 9476 & 184399 & -4918.82 & 1525 & 4 \\ \hline 
  1IGT & 20176 & 426588 & -12261.13 & 4239 & 5 \\ \hline 
\end{tabular}
\caption{\label{table:size} Solvation energy, time to solution using \texttt{GPU}, and number of self-consistent iterations for different sized biomolecules.}
\end{table}

\begin{table}
\begin{tabular}{|c|c|c c c|c|c|c|}\hline
\textbf{Force} & $\epsilon_\text{p}$ &  \multicolumn{3}{c|}{\textbf{G$_\text{solv}$ [kcal/mol]}} & \textbf{$\Delta$G$_\text{bind,vac}$} & \textbf{$\Delta\Delta$G$_\text{bind}$} & \textbf{$\Delta$G$_\text{bind,diss}$}  \\ 
\textbf{field}   & & \textbf{Complex} & \textbf{HIV+CD4} & \textbf{Antibody} & \textbf{[kcal/mol]} & \textbf{[kcal/mol]} & \textbf{[kcal/mol]} \\ \hline
   \texttt{AMBER}  & 1 & -8754.31  & -5181.56  & -4128.96 & -508.50 & 556.21 & 47.71 \\ 
                   & 2 & -4291.84  & -2543.03  & -2023.35 & -254.84 & 274.54 & 19.70   \\ 
                   & 4 & -2062.65  & -1225.64  & -972.00  & -127.13 & 135.00 & 7.87    \\
                   & 6 & -1322.33  & -787.81   & -622.69  & -85.00 & 88.17 & 3.17  \\ \hline
   \texttt{CHARMM} & 1 & -8799.66  & -5187.31  & -4169.20 & -519.70 & 556.85 & 37.15   \\ 
                   & 2 & -4313.50  & -2546.07  & -2043.32 & -259.84 & 275.89 & 16.05  \\
                   & 4 & -2073.57  & -1227.20  & -981.84  & -129.93 & 135.5 & 5.57   \\
                   & 6 & -1329.51  & -788.98   & -629.17  & -86.61 & 88.64 & 2.03   \\ \hline
   \texttt{PARSE}  & 1 & -12151.82 & -6926.11  & -5839.22 & -552.72 & 613.51 & 60.78  \\
                   & 2 & -5955.71  & -3395.56  & -2862.21 & -276.35 & 302.06 & 25.71  \\
                   & 4 & -8260.06  & -1632.20  & -1376.24 & -138.18 & 148.38 & 10.2  \\
                   & 6 & -1834.25  & -1047.78  & -882.49  & -92.12 & 96.02 & 3.9  \\ \hline
   \texttt{AMOEBA} & 1 & -8511.01  & -4846.20  & -3547.57 & 152.25 & -117.24 & 35.01 \\ 
   w/$\alpha=0$    & 1 & -9724.01  & -5541.78  & -4121.54 & 113.32 & -60.69 & 52.63  \\ \hline 
\end{tabular}
\caption{\label{table:bind} Electrostatic component of binding energy of \texttt{1GC1} computed with the thermodynamic cycle in Figure \ref{fig:thermo_cycle}.}
\end{table}

\end{document}